\documentclass[conference]{IEEEtran}
\IEEEoverridecommandlockouts
\usepackage{cite}
\usepackage{amsmath,amssymb,amsfonts}
\usepackage{graphicx}
\usepackage{textcomp}
\usepackage{xcolor}
\usepackage{algorithm}
\usepackage{algorithmicx}
\usepackage{algpseudocode}
\usepackage{subcaption}
\usepackage{booktabs}
\usepackage{tikz}
\newcommand*\circled[1]{\tikz[baseline=(char.base)]{
            \node[shape=circle,draw,inner sep=0.7pt] (char) {#1};}}
\usepackage[hyphens]{url}

\def\BibTeX{{\rm B\kern-.05em{\sc i\kern-.025em b}\kern-.08em
    T\kern-.1667em\lower.7ex\hbox{E}\kern-.125emX}}
    
\begin{document}

\title{Secure and Efficient Trajectory-Based Contact Tracing using Trusted Hardware}

\author{
    \IEEEauthorblockN{Fumiyuki Kato}
    \IEEEauthorblockA{\textit{Department of informatics}\\
    \textit{Kyoto University}\\
    Kyoto, Japan \\
    fumiyuki@db.soc.i.kyoto-u.ac.jp}
    \and
    \IEEEauthorblockN{Yang Cao}
    \IEEEauthorblockA{\textit{Department of informatics}\\
    \textit{Kyoto University}\\
    Kyoto, Japan \\
    yang@i.kyoto-u.ac.jp}
    \and
    \IEEEauthorblockN{Masatoshi Yoshikawa}
    \IEEEauthorblockA{\textit{Department of informatics}\\
    \textit{Kyoto University}\\
    Kyoto, Japan \\
    yoshikawa@i.kyoto-u.ac.jp}
}

\maketitle

\begin{abstract}
The COVID-19 pandemic has prompted technological measures to control the spread of the disease. 
Private contact tracing (PCT) is one of the promising techniques for the purpose.
However, the recently proposed Bluetooth-based PCT has several limitations in terms of functionality and flexibility.
The existing systems are only able to detect \textit{direct contact} (i.e., human-human contact), but cannot detect \textit{indirect contact} (i.e., human-object, such as the disease transmission through surface).
Moreover, the rule of risky contact cannot be flexibly changed with the environmental situation and the nature of the virus.
In this paper, we propose a secure and efficient trajectory-based PCT system using trusted hardware.
We formalize trajectory-based PCT as a generalization of the well-studied Private Set Intersection (PSI), which is mostly based on cryptographic primitives and thus insufficient.
We solve the problem by leveraging trusted hardware such as Intel SGX and designing a novel algorithm to achieve a secure, efficient and flexible PCT system.
Our experiments on real-world data show that the proposed system can achieve high performance and scalability.
Specifically, our system (one single machine with Intel SGX) can process thousands of queries on 100 million records of trajectory data in a few seconds.
\end{abstract}


\section{Introduction}
\label{sec:1}
Since the beginning of 2020, the emergence of the COVID-19 is causing a worldwide pandemic.
Many governments and companies are developing various measures and technologies to prevent the spread of the virus \cite{wang2020response, qin2020dysregulation, salathe2020covid, cho2020contact}.
Currently, we do not have any effective treatment and vaccine available, and it may take one or a few years before the advent of them to bring the situation under control.
At present, contact tracing is expected as a powerful countermeasure for controlling the spread of infection. 
The effectiveness of contact tracing has already been shown by several previous studies \cite{ferretti2020quantifying, reichert2020privacy, tang2020privacy, brack2020decentralized}.
However, conducting an effective contact tracing often need to collect citizens' personal information like locations \cite{privatedata} or telephone number \cite{singepore}, which raises ethical issues and serious privacy violations \cite{jensen2009location}.
Therefore, private contact tracing (PCT) is urgently needed.

Recently, Bluetooth-based private contact tracing has been intensively studied \cite{troncoso2020decentralized, trieu2020epione, rivest2020pact, cen, gvili2020security}. 
Decentralized Privacy-Preserving Proximity Tracing (DP3T) \cite{troncoso2020decentralized}, which is an open protocol for PCT using Bluetooth Low Energy beacons, is already being used in applications developed in Europe.
To strongly protect users' privacy, it uses only contact (proximity) history detected by the Bluetooth Low Energy beacons.
In DP3T, the base mechanism is that the applications use a Bluetooth of the smartphone to broadcast a random ID that does not include sensitive information such as the user's identity or location, and the nearby smartphone devices receive and store the data for a limited time.
Users who are then discovered to be infected with the coronavirus send a report to the server that includes the random IDs they have generated.
Meanwhile, the user routinely checks to see if the random IDs received from devices they have contacted in the past have been uploaded to the server.
Additionally, there are similar methods adopting decentralized architecture such as Epione \cite{trieu2020epione}, PACT protocol \cite{rivest2020pact}, CEN \cite{cen} and Google and Apple specification \cite{gvili2020security}.

However, the Bluetooth-based private contract tracing has several limitations in terms of functionality and flexibility.
First, Bluetooth-based PCT only detects \textit{direct contact} (i.e., human-human contact), but cannot detect \textit{indirect contact} (i.e., human-object, such as the disease transmission through surface).
The Centers for Disease Control and Prevention (CDC) in US has shown that it is possible that a person can get COVID-19 by touching a surface or object that has the virus on it and then touching their own mouth, nose, or eyes \cite{CDC} --- even she does not have direct contact with COVID-19 patients.
Second, the Bluetooth-based PCT lacks flexibility in terms of determining the rule of ``risky contact".
Essentially, the rule of risky contact in the Bluetooth-based PCT is hard-wired into the Bluetooth device since the risky contact is implicitly defined as two devices are in close proximity to each others' signal range.
In practice, whether or not it is a risky contact varies with the environmental situation and the nature of the virus.
In fact, the rules of risky contact in COVID-19 has been updating along with the understanding of the virus \cite{EVOLVE}. 
For example, in the beginning of the pandemic, professionals believed that transmission only takes place through direct human-human contact; however, recently, it was argued that airborne transmission should be taken into account \cite{AIRBORNE}.

\begin{figure}[t]
  \includegraphics[width=\linewidth]{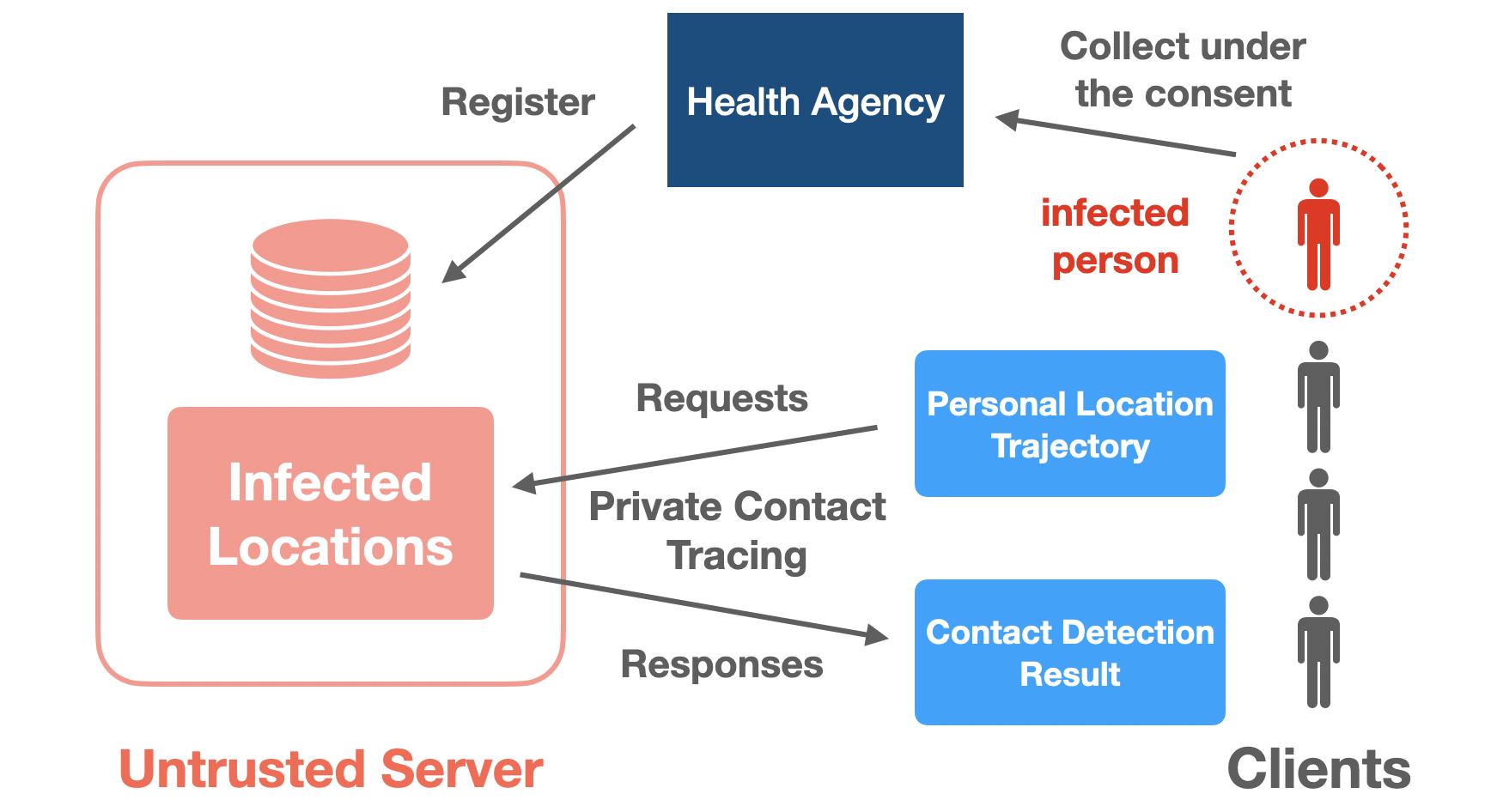}
  \caption{Trajectory-based PCT overview.}
  \label{fig:over_view}
\end{figure}

In this work, we propose a secure and efficient \textit{trajectory-based private contact tracing} to enable both direct and indirect contact tracing.
By comparing the trajectory data between a user and the infected patients, we can check whether the user visit the ``infected locations'' or not.
The rule of risky contact can be flexibly defined according to the condition of a location and the nature of the virus.
We list four requirements for secure and efficient trajectory-based private contact tracing.

\begin{enumerate}
  \item {\itshape Security}: A client's trajectory data  must be protected from the server and any other clients. 
  On the other hand, nothing about the server side data is disclosed to the client except the query result. 
  \item {\itshape Efficiency}: The query throughput that can be handled by the central server is crucial.
  \item {\itshape Flexibility}: The rule of risky contact should be flexibly changeable.
  \item {\itshape Correctness}: The server must return the true result because the result is sensitive and can significantly affect the users.
\end{enumerate}

\noindent
As shown in Figure \ref{fig:over_view}, we assume that the health agency
registers location data of the confirmed COVID-19 patients (these data are encrypted or released under the consent of the patients) to a server which is untrusted by clients.
The server receives queries and encrypted personal trajectories from clients and returns a boolean value of whether or not there is a risky contact.

The problem of trajectory-based PCT is similar to the well-studied problem of Private Set Intersection (PSI), which ensures that two (or more) parties collaboratively calculate the intersection of their private sets, while nothing about the private data will be disclosed to the other party except the existence information of intersection or the result.
However, existing techniques for PSI, mostly based on cryptographic primitives, cannot achieve all of the above-mentioned requirements.
The cryptography-based PSI approach such as MPC techniques or Oblivious RAM (ORAM) has limitation in terms of efficiency, there are still performance problems \cite{narayanan2011location,kiss2017private, reichert2020privacy} in medium or large workload.
This is mainly due to the heavy use of time-consuming cryptographic primitives.
Recently, secure hardware (such as Intel SGX or ARM TrustZone) based approaches have received increasing  attentions.
It enables Trusted Execution Environment (TEE) \cite{sabt2015trusted, costan2016intel}, which is used for speeding up secure computation on untrusted parties.
Tamrakar et al.  \cite{tamrakar2017circle} proposed the first efficient TEE-based PSI. 
However, it does not satisfy our requirement of correctness since it introduces a non-zero false positive rate because of using probabilistic data structures.
In addition, flexibility has not been achieved.

Our contributions in this paper are three-folded.
First, we formulate the problem of trajectory-based PCT. 
We show that our problem (due to the requirement of flexibility) is a generalization of well-studied Private Proximity Testing \cite{narayanan2011location} and Private Set Intersection (PSI).
Our formulation is parametrized for both time and space and can be used in general settings.
Second, we propose a TEE-based system architecture and efficient algorithms for trajectory-based PCT.
Besides satisfying the above-mentioned requirements, a challenge in designing TEE-based algorithm is the constraint of the secure memory (i.e., enclave) on secure hardware.
We solve these problems by designing a novel trajectory data encoding method, which enables algorithmic flexibility and furthermore provides efficient compression using finite state automaton which also provides deterministic and fast search performance for high speed PSI.
Third, we implement the proposed system on Intel SGX and open source the code in GitHub\footnote{https://github.com/ylab-public/PCT}.
Our experiments on real-world datasets show that the proposed system is efficient and effective in practical scenarios.
Specifically, the proposed encoding and data structure compresses the actual trajectory data to one-sixth the size of Hash Table with the same performance, and as a result, the total runtime is significantly reduced.
Moreover, we show that our system implemented on a single machine equipped with SGX can handle thousands of queries on 100 million records of trajectory data in a few seconds.

\vspace{3px}
\noindent
{\bfseries Outline.} 
In Section \ref{sec:2}, we show some features of Intel SGX which are essential for understanding our architecture.
In Section \ref{sec:3}, we describe problem statement and formulate PCT problem.
In Section \ref{sec:4}, we discuss the details of our architecture, algorithm and trajectory-based data compression.
In Section \ref{sec:5}, we show experimental results and evaluation, and then discuss future work in short in section \ref{sec:6}.
In Section \ref{sec:7}, we show related works.
Finally, we conclude in Section \ref{sec:8}.

\section{Intel SGX}
\label{sec:2}
In this section, we introduce the secure hardware used in this paper for the ease of understanding our system.

Intel SGX\cite{costan2016intel} is the extended instruction set of Intel x86 processors, which enables the creation of an isolated trusted execution environment, called {\itshape enclave}. 
In addition to powerful server machines, it is installed on some PCs.
It is also available on some public cloud platforms, Azure Confidential Computing, Alibaba Cloud, IBM Cloud, and so on.
We show a brief overview of SGX in the following paragraphs.

{\itshape Enclave} resides protected memory region, called Enclave Page Cache (EPC), in which all programs and data can be unencrypted and fast processed while they transparently encrypted outside CPU package by memory encryption engine using a secret key which only processor hardware can access.
In other words, SGX adopts a model that considers the CPU package as a trust boundary and everything outside as untrusted.
In this trusted space, accesses from any untrusted application, including OS/Hypervisor, are prohibited by the CPU, protecting the confidentiality and integrity of the program and data inside {\itshape enclave}.
This hiding process is transparent to the software and it is extraordinarily faster than the conventional advanced cryptography-based method in secure computation. 

While the capability to perform such fast, confidential computation has attracted a lot of attention, various research targets its vulnerability. 
Especially side channel attacks\cite{chen2019sgxpectre, gotzfried2017cache} and the effective countermeasures\cite{orenbach2020autarky} have been active recently.
In this work, we do not consider side-channel attacks and assume that the hardware is secure.

\noindent
{\bfseries Memory size limitation.}
A challenge in designing algorithms for Intel SGX is the size constraint of EPC.
The maximum size of EPC is limited to 128MB, including 32MB meta-data for secure management (or 256MB including 64MB meta-data in recent Intel hi-end processor\cite{hiendintel}).
This limitation actually may be improved gradually, but it will continue to be a problem for hardware and memory securing performance.
Suppose memory is allocated beyond this memory size constraint. 
In that case, SGX with Linux allows paging with special encryption. 
However, many studies have shown that the performance is hugely degraded by severe overhead \cite{gueron2016memory, gjerdrum2017performance, taassori2018vault}, which is derived from a requirement to preserve confidentiality and integrity even outside enclave.
Therefore, it is necessary to design an efficient algorithm that works within SGX. 

\noindent
{\bfseries Attestation.} 
SGX supports two types of attestations, local and remote, which can verify the correct initial state and genuineness of the trusted environment of the enclave from outside.
In this section, we focus on the remote attestation (RA)\cite{costan2016intel}.
We can request RA to the enclave and receive a report with measurements (e.g., MRENCLAVE and MRSIGNER) based on the hash of the initial enclave state, which can identify the programs, complete memory layout, and even builder's key information, and nobody can tamper with this measurement.
Intel Enhanced Privacy ID\cite{costan2016intel} signs this measurement, and the Intel Attestation Service can verify the correctness of the signature as a role of trusted third party.
In addition to verifying the SGX environment, secure key exchange, such as ECDSA, between the enclave and remote client is performed within this RA protocol.
Therefore, after that, we can communicate over a secure channel with a remote enclave by standard encryption and, finally, we can safely perform a confidential calculation in the remote enclave.

\section{Problem Formulation}
\label{sec:3}
In this section, we first introduce the scenario of trajectory-based private contact tracing and then we formulate our problem based on the well-studied private proximity testing.

\subsection{Our Scenario}
\label{sec:3-1}
In our scenario, we assume that trajectory-based PCT is used to prevent the spread of COVID-19.
We consider a centralized architecture that stores the trajectory data of infected patients on a central server and accepts PCT requests from users with their trajectory data.
In practice, the infected patients' trajectories can be received in bulk from public institutions such as a government or health agency.

In the operation of the system, based on the incubation period of the virus, the server always keeps the trajectory data of the infected patients for the past 14 days (up to 21 days). 
It periodically updates in batches (e.g., once a day), adding and deleting data. 
The server transforms the trajectory data to an appropriate structure in advance and is always ready to accept PCT requests from clients.
The client also sends encrypted trajectory data for the past 14 days as a PCT request and the server performs the contact detection and then returns the results to the client.
The results are time-stamped and signed in SGX so that they can be verified by a third party, allowing clients to use them in various agencies and events to show that the risk of infection is low.

Finally, the threat model is a malicious server with privileges for software/hardware, and honest clients.
The client can only trust the CPU package equipped with SGX used on the server.

\subsection{Problem Statement}
\label{sec:3-2}

\noindent
{\bfseries Trajectory-based PCT.}
The trajectory-based PCT protocol is an asymmetric protocol between client and server.
When a client wants to know the contact with server $v$'s trajectories, this protocol returns 1 or 0 to the client, depending on the result, and does not disclose the information of the client to the server.
In the use case for infections, each client has a set of trajectory data for one person, and the server has trajectory data for a lot of infected patients.
In conventional private proximity testing \cite{narayanan2011location}, when two people, user $u$ and user $v$, have geographic data $X_i=l^{(i)}_t$ of time t. When $u$ execute the protocol, $u$ can get the results as follows,

$$
\begin{cases}
    1 \;\; (\,|| l^{(u)}_t - l^{(v)}_t || \le \delta\,) \\
    0 \;\; (\, others \,)
\end{cases}
$$

\noindent
where $\delta$ is proximity threshold.
After that, $v$ does not learn any information about $X_u$ and $u$ does not learn information except $|| l^{(u)}_t - l^{(v)}_t || \le \delta $.
In the simplest form, trajectory-based PCT can be represented as an extension of such a formulation.
For contact tracings, a contact can be determined according to human time-series tracking data.
We can do that based on private proximity testing by extending a single geographic data to time-series trajectory data.
A threshold can also be extended to two-dimensional thresholds to check spatial-temporal proximity.
PCT allows for the capture of indirect contact by confirming that the patients are in the same place within a specific period.
Therefore, we can formulate as follows, denoting trajectory data of user $k$ as $X_k=(x^{(k)}_1=(t^{(k)}_1,l^{(k)}_1),...,x^{(k)}_n=(t^{(k)}_n,l^{(k)}_n))$, $u$ can get the result of contact with $v$,

\begin{eqnarray}
\begin{cases}
    1 \;\; (\,x^{(u)}_i \in X_{u}\,, \;x^{(v)}_j \in X_{v} \,s.t.\\
    \;\;\;\;\;\;\;\; ||l^{(u)}_i - l^{(v)}_j|| \le \delta_{geo} \,\; \mathrm{and} \,\; ||t^{(u)}_i - t^{(v)}_j|| \le \delta_{time} \,) \\
    0 \;\; (\, others \,)
\end{cases}
\end{eqnarray}

\noindent
where $\delta_{geo}$ is spatial and $\delta_{time}$ is temporal proximity threshold. 
Furthermore, the server does not learn any information about $X_u$, and $u$ can get only 1 or 0 about $X_v$ in this protocol.

This problem is similar to set intersection problem between $X_u$ and $X_v$.
For all trajectory data, if we discretize the geographic data using squares of length $\frac{\theta_{geo}}{\sqrt{2}}$ on one side and discretize the time data at intervals of ${\theta_{time}}$, for all the set intersection $X_u \cap X_v$ satisfies the formula (1).
Thus, PSI can be used as an approximation to PCT.
Note that the discretization has the limitation that it obscures the accuracy around the boundary of discretization, compared to (1).

\vspace{5px}
\noindent
{\bfseries Efficiency.}
Trajectory-based PCT requires efficiency in several aspects.
The first is response throughput since the server will always be exposed to requests from a large number of users.
This number cannot be a problem if a peer-to-peer protocol is employed like previous decentralized PCT architecture, while it can be significant workloads in such a centralized one.
The second is the bandwidth. Since the protocol is applied to many users, it is necessary to reduce the bandwidth for communication efficiency.
In other words, protocols that require communicating a lot of information to edge devices should be avoided.
This bandwidth can often be a major cost for cryptographic methods that require preprocessing \cite{kiss2017private} before contact judgment algorithm.
The third is scalability.
For instance, in COVID-19, the size of the infected patients's data and the user's size may increase in the event of a spread of infection.
The efficiency requirements depends entirely on the context in which PCT is deployed and is determined by the number of users, frequency of use, number of data, etc.

\vspace{5px}
\noindent
{\bfseries Security.}
Considering the threat model in trajectory-based PCT, the most prominent and necessary model which we should defend against is the malicious server.
Malicious server tries to obtain information illegally without the constraint of following the protocol.
In a typical model, the server runs in the untrusted software and hardware environment, and the privileged attacker has full control over the OS and/or Hypervisor, memory hardware units, and packet monitoring in the network.
Under this model, it is essentially necessary to have cryptographic indistinguishability in the PCT processing on the server and delivering on the network to protect user's privacy, because the adversary can monitor the processing of raw data.
On the contrary, we can assume that an honest-but-curious client follows the protocol but tries to obtain private information from the server.
They can perform a brute-force attack to partially reveal trajectory data on the server-side if the trajectory data space does not have large entropy.
The malicious client can send more extensive background knowledge and crafted data, but the countermeasures are basically the same.
By making the value returned to the user binary, we can reduce the amount of information leaked and increase the attacker's cost by authenticating each user and setting a limit on the number of requests.
In order to seek more complete protection, we can adopt probabilistic indistinguishability definitions like differential privacy.

\vspace{5px}
\noindent
{\bfseries Flexibility and Correctness.}
Flexibility, expressed a little more formally, is the requirement that $(\delta_{geo}, \delta_{time})$ be parameters in the system.
For example, these parameters need to be changed to minimal values if it is found after the system released that only direct contact is needed to be captured because of the virus's capacity for transmission.
Correctness does not allow the PCT to return any probabilistic answer, highly dependent on the domain being used.

\section{Proposed System}
\label{sec:4}
We first introduce an overview of the system, and then analyze the proposed system to show how the system meets the requirements.
Next, we show how to turn the trajectory data into a compressed dictionary representation, and finally, we describe the overall algorithm description and computational cost.
Table \ref{tab:params} shows the symbols and parameters that are used in the following sections.

\begin{table}[t]
\centering
\caption{Symbols and parameters.}

\begin{tabular}{cl}
    \toprule
    Symbols & Explanation\\
    \midrule
    $(t,l)$ & tuple of time and location, one row of trajectory data \\
    $D$ & raw trajectory data of infected people \\
    $N_{D}$ & number of chunks of central data \\
    $N_{C}$ & number of clients \\
    $c_{i} \in \mathbf{C}$ & a client $i(\in \{1,...,N_{C}\})$ and all clients set \\
    $R$ & mapped $D$, array of efficient chunks $(=(r_1,...,r_{N_D}))$\\
    $r_{i}$ &  $i$-th chunked data of $R$, efficient representation (e.g., FSA) \\
    $q_{i}$ & client $i$'s query data (raw trajectory data) \\
    $Q$ & merged and mapped $N_{C}$ query data (e.g., unique array)\\
    $N_Q$ & unique size of $Q$ \\
    $\theta$ & parameter of PCT, $\theta=(\theta_t, \theta_l)$ \\
    \bottomrule
\end{tabular}

\label{tab:params}
\vspace{-1em}
\end{table}

\subsection{System Overview}
\label{sec:4-1}

\begin{figure}[t]
  \includegraphics[width=\linewidth]{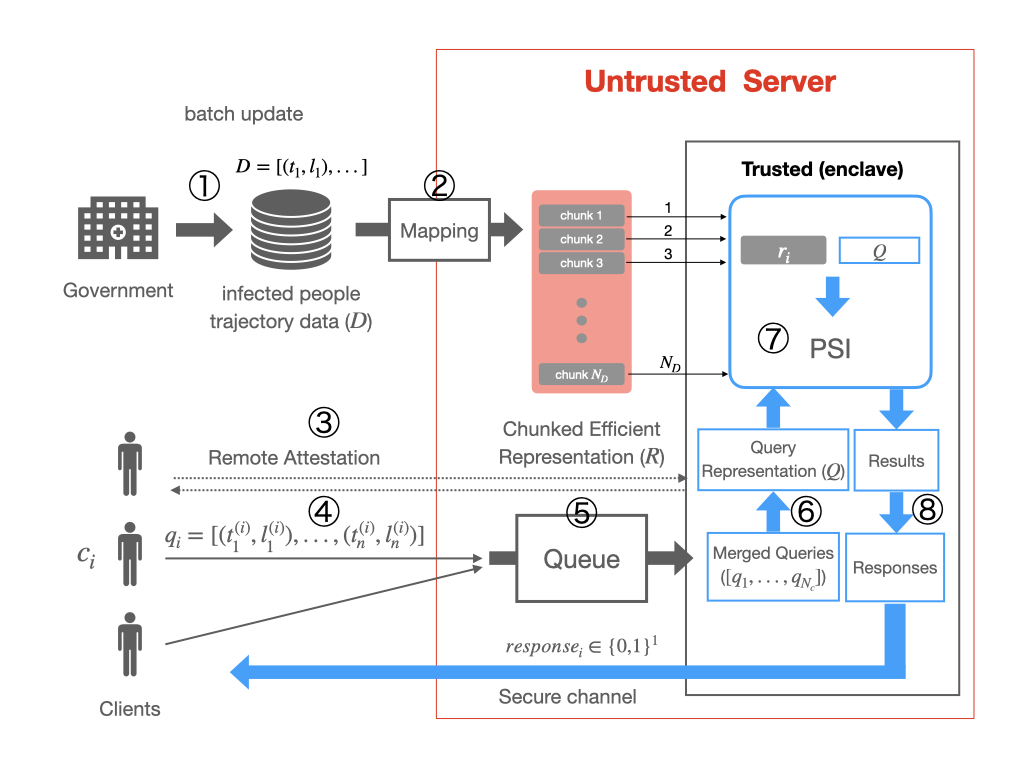}
  \caption{Architecture overview: the circle numbers corresponds steps shown in {\itshape System Description}.}
  \label{fig:carousel}
\end{figure}

Figure \ref{fig:carousel} shows the overview of our architecture using trusted enclave.
Our method consists of several steps, including the transformation of data maintained on the server side and the transformation of data sent from the client side, as follows. First part, we describe the data of infected patients on the server side.

\noindent
\circled{1} : (Update master data) Government updates the infected patients data $D$ in batch processing.
$D$ is in the raw form of trajectory data $D=[(t_1,l_1),...,(t_{n_R},l_{n_R})]$.

\noindent
\circled{2} : (Mapping) We map from the raw data format $D$ to efficient dictionary representation $R$ with function $mapToChunkedDictionary(D, \theta)$.
$R$ consists of $N_D$ chunks $r_i$ $(i=1,...,N_D)$, where each chunk fits in the enclave memory.
The function takes the parameter of PCT $\theta$ and use it as the granularity parameter for the chunks.
Here, each chunk $r_i$ is encoded as a 1D string suitable for PSI.
The encoding scheme is described in the next section.
Step 2 is executed in the same batch processing as step 1, with new data being added and expired data being deleted at that time.

The next part is the processing of queries from clients.

\noindent
\circled{3} : (Remote attestation) Client verify the remote enclave through remote attestation protocol before sending the request to the server.
The client can confirm that the enclave has not been tampered with and then securely exchange keys with the enclave.
Thereafter, the key is used to encrypt the data, which enables secret communication to the remote enclave through a secure channel.

\noindent
\circled{4} : (Request) Many clients send PCT requests to the server.
In the figure, $c_i$ sends $q_i$ as a parameter of the query which contains her trajectory data.
$q_i$ is encrypted in all the untrusted area and is only visible in the verified enclave.

\noindent
\circled{5} : (Queuing) Until a certain number ($N_c$) of requests are accumulated, $q_i$ is queued outside the enclave and they are passed to enclave all together by $loadToEnclave$ function.
This function is actually implemented by the so-called {\itshape ECALL} in SGX programming.

\noindent
\circled{6} : (Mapping) Inside the enclave, a number of $q_i$ are grouped together and mapped to query representation $Q$ using $mapToUniqueArray$.
The function also takes query data and a granularity parameter $\theta$ the same as step 2.
We intend to convert the data structure suitable for PSI, basically, $Q$ is represented as a unique array.

\noindent
\circled{7} : (Contact detection): The chunked data $r_i$ is imported into enclave one by one, and we computes the set intersection of $r_i$ and $Q$ in enclave.
This can be done by checking the string-based match with the transformation in step 2.
The results are stored together.

\noindent
\circled{8} : (Response construction) After the iterations for all the chunks are done, responses for all clients are constructed from the results and complete query data $q_i$ $(i=1,...,N_C)$ inside trusted enclave by $constructResponses$.
This can be done by simply checking that wheather each query has data of the results.
Finally, it returns the encrypted result through the secure channel to each client.

\subsection{System Analysis}
\label{sec:4-2}

\noindent
{\bfseries Efficiency.}
The most important point is the use of SGX for the private calculation in step 7.
SGX allows software to perform secret computations transparently and eliminates the need for complicated and time-consuming cryptographic techniques to perform PSI.
The computational overhead is small and the overall speed is improved dramatically.
On the other hand, step 2 is important to overcome the weakness of SGX.
The chunking, $R$ into $r_i$ $(i=1,...,N_D)$, avoids serious paging overhead caused from severe memory constraint of SGX even when the infected patients' data become too large to fit into the enclave.
Moreover, the mapping in step 2 involves data transformation along with chunking.
The raw data representation of trajectories is encoded into a string, and then converted into the memory-efficient dictionary representation.
As we show in the following section, the data structure of $r_i$ achieves memory savings by using finite state automaton, which also provides fast search.
Finally, step 5 and step 6 are to improve the throughput of the query processing.
Reading the chunked data $r_i$, as in Step 7, is costly due to the $N_D$ iteration, and doing this for every query can have a large overhead.
This overhead can be mitigated by multiplexing the query for efficiency.

\noindent
{\bfseries Security.}
While using SGX enables efficient query processing, it also provides high security.
It ensures security against malicious servers with any software privileges and prevents client private data from being exposed in any untrusted areas.

In our architecture, data sent by clients according to protocol of SGX is hidden by encryption outside of enclave.
To be precise, in step 5, $q_i$ is decrypted using shared key exchanged through Remote Attestation in step 3 after it is loaded into the enclave, and in step 8, it is encrypted before sending it out of the enclave. 
Besides, because the enclave environment and the programs running in it are verified by a prior remote attestation from the client, the client can always reject the request if there is non-private processing going on in the enclave.
Therefore, cryptographically strong security for the client's privacy from any external attacker is ensured in case using proper encryption and without software vulnerability.

\noindent
{\bfseries Flexibility.}
The requirement of flexibility is addressed by the mapping at step 2 and step 6.
These mappings are responsible for what granularity the trajectory data is quantized and encoded.
Mapping means the transformation of raw data to a set of strings, that is encoding, and this mapping is parameterized with granularity of $\theta$.
The parameter is shared between the server(step 2) and clients(step 6).
We can achieve flexibility in our PSI-based methods by this parameterization.
In these steps, while step 2 is a batch processing, step 6 is an online processing, so it is more efficient to send preprocessed data to the server side to distribute the processing load to clients.

\noindent
{\bfseries Correctness.}
This is achieved without any probabilistic guarantees or extra systems, as it is transformed into a deterministic data structure, FSA, by step 2 mapping.
Using this, after we run the PSI in step 7, the results are definitely correct.

\subsection{Trajectory-based data compression}
\label{sec:4-3}
The challenge we need to solve here is how to perform efficient query processing under the severe memory constraints of SGX in step 7.
In particular, we have to carefully consider the dictionary representation $R$ ($=(r_1,...r_{N_D})$) obtained by the mapping in step 2.
$R$ is the most important data structure for PSI and should have the following requirements.
First, it should have memory-efficient data structure storing trajectory data to overcome severe memory constraints.
Second, it should have fast search performance, because in large scale PSI, it needs to process a lot of checks if it includes target or not.
Third, it should provides deterministic search method for correct PSI.
Another challenge is how to represent trajectory data.
While we need to encode different trajectory data into a unique string to perform PSI, we show that encoding considering the nature of trajectory data helps efficient compression.

Standard dictionary representations do not match our requirements.
A well-known data structure for dictionary representation is Hash Table, and we use Hash Table as a baseline for our experiments.
The Hash Table is ideally supporting the $\mathcal{O}(1)$ key-based search.
While Hash Table can provide desirable search performance and deterministic search, they fail to satisfy the first requirement due to their linear size increase with the number of data it store.
A smaller data structure is preferable in our setting because overhead of SGX constraint are heavy.
While probabilistic data structures such as Bloom Filter provides the same speed of search performance as Hash Table and superior memory efficiency, they do not satisfy the third requirement because they cannot provides deterministic search.

Our proposed method to obtain the desired dictionary representation is a combination of encoding and data structure.
Roughly speaking, the encoding process transforms trajectory data into highly similar strings representation, and then utilize the similarity to create a compressed dictionary representation using the Finite State Automaton (FSA).
FSA is a deterministic finite state acceptor and can be cyclic and permits the sharing of both prefixes and suffixes among the same nodes (described later section).
Therefore, similar strings can be stored efficiently.

\noindent
{\bfseries Trajectory-based encoding.}
First, we introduce trajectory-based encoding.
The first property that our encoding must satisfy is that there is injective function between, discretized and different trajectory data and unique strings.
Obviously, if this property is not satisfied, PSI cannot be performed correctly.
Another desired property is that the string has many similarities in suffixes and prefixes which is because of FSA.

\begin{figure}[t]
  \setlength{\belowcaptionskip}{-10pt}
  \includegraphics[width=\linewidth]{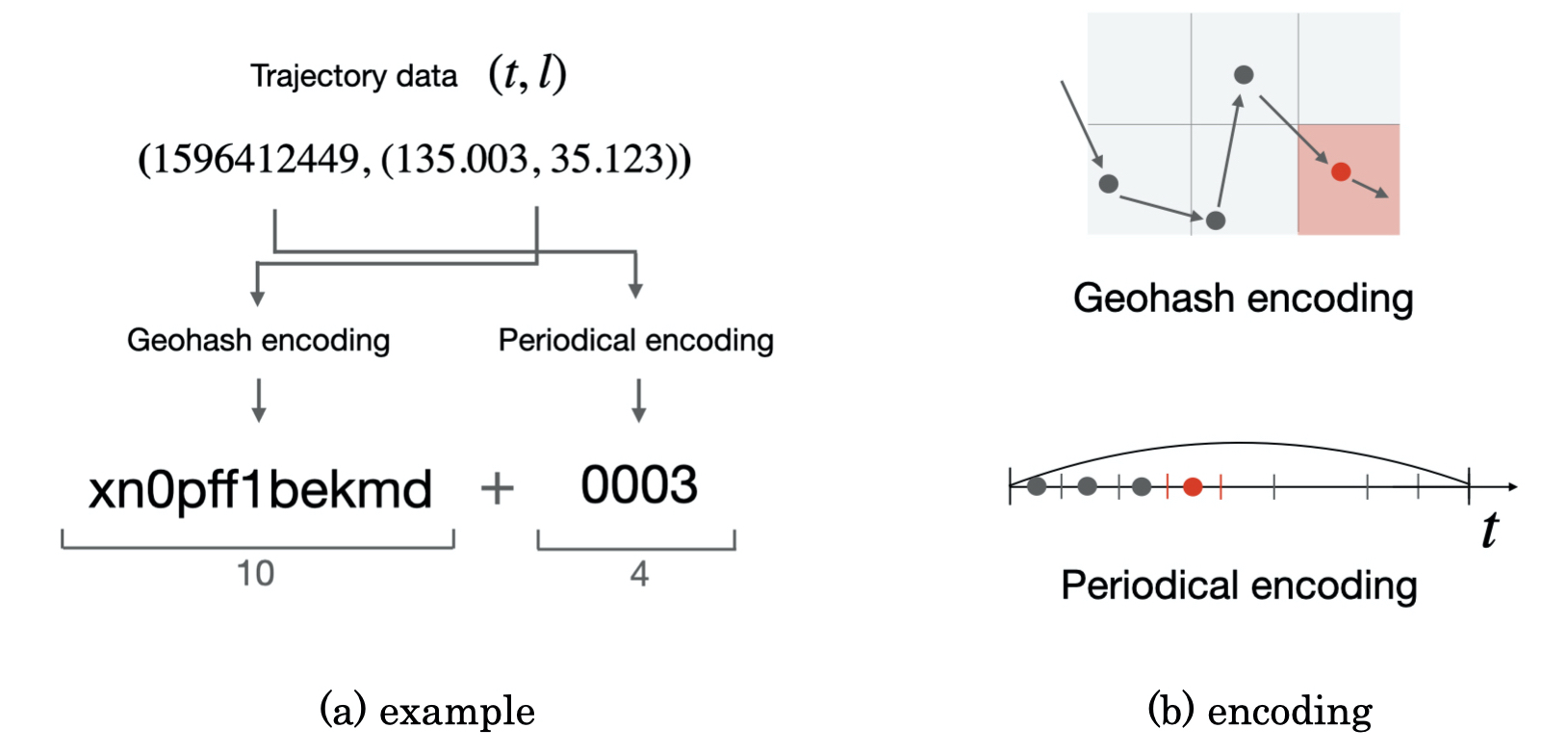}
  \caption{Encoding design: transforming trajectory data into string. Figure (a) shows an example encoding. Figure (b) shows visualization of our encoding method.}
  \label{fig:encode}
\end{figure}

We show our encoding that satisfies both properties.
The trajectory data $X$ consists of an array of tuples of temporal data such as UNIX epoch and geographical data such as tuple of latitude and longitude, as follows.
\begin{equation*}
\begin{gathered}
X = [(t_1,l_1),...,(t_n, l_n)] \\
t_k \in \mbox{time (e.g. UNIX epoch)} \\
l_k \in \mbox{coordinate (e.g. (latitude,\,longtitude)})
\end{gathered}
\end{equation*}
We encode $t$ and $l$ separately and merge them in series, as shown in Figure \ref{fig:encode}(a).
For $l$, there is a method of mapping two-dimensional coordinate data to a string; a typical example is Geohash \cite{geohash}.
Geohash is an encoding method for representing coordinate information in terms of latitude and longitude as Base32 of any length according to the desired grain size, recursively dividing the area on the map into squares.
In Geohash, the strings obtained from the trajectory data are likely to be similar in suffix, because the data at close distances can have the same strings in the order from the beginning.
For $t$, the most naive way to encode is to use the time representation in seconds as a string, but it can be possible to quantize it to a shorter length within a specific period to match the survival of the virus.
This {\itshape Periodical encoding} divides the period into $n$ segments of equal size, allocates numbers from $0$ to $n-1$, and then encodes the numbers themselves.
This encoding generates similar strings for all trajectory data of the same period.
And then, simple merging leads to frequent occurrences of structures similar to prefix and suffix, which can be compressed more efficiently by the FSA.
It is also clear that the encoded string is different for different trajectory data.
Thus, our encoding satisfies the encoding requirements mentioned above.

As an example, Figure \ref{fig:encode} (a) shows a row of trajectory data is mapped to a character string of length 14.
(b) intuitively describes two types of encoding: Geohash encoding, which discretize the trajectory data into blocks of squares on the map, and Periodical encoding, which divides the trajectory data into equal time intervals and represents the number.

As an additional note on encoding, for the flexibility, an important aspect is that both of these encodings can be parameterized for granularity.
The Geohash encoding is controlled for granularity with the parameter $\theta_t$. 
For example, when the number of digits of Geohash corresponds to $\theta_t$, $\theta_t = 10$ means a rectangular block with a granularity of 50 cm or less.
Similarly, the Periodical encoding determines the segmentation granularity by parameter $\theta_t$.
Therefore, when we would like to change the PCT conditions, we only need to change the $\theta$ in encoding.

\noindent
{\bfseries Desired data structure.}
In this section, we explain that FSA satisfies our three requirements mentioned in \ref{sec:4-3}.
Trie \cite{trie1960} is a deterministic acyclic finite state acceptor and can store and compress string data, sharing string prefixes from the tree structure's roots.
They also provide a fast string search as a dictionary in proportion to the maximum depth. 
However, the search cost can be $\mathcal{O}(1)$ if the maximum length is small, which is asymptotically equivalent to Hash Table and may be advantageous because it does not need computing hash functions.
Thus, it basically meets our requirements, but actually the compression capability is somewhat unclear.
FSA (Finite State Automaton) has basically similar capabilities as Trie, but it can permit the sharing of both prefixes and suffixes.
\cite{Daciuk2011SmallerRO} shows its effectiveness by extensive experiments.
Therefore, we increase the compression efficiency by introducing FSA, and this data structure satisfies our requirements.

\begin{figure}[t]
  \includegraphics[width=\linewidth]{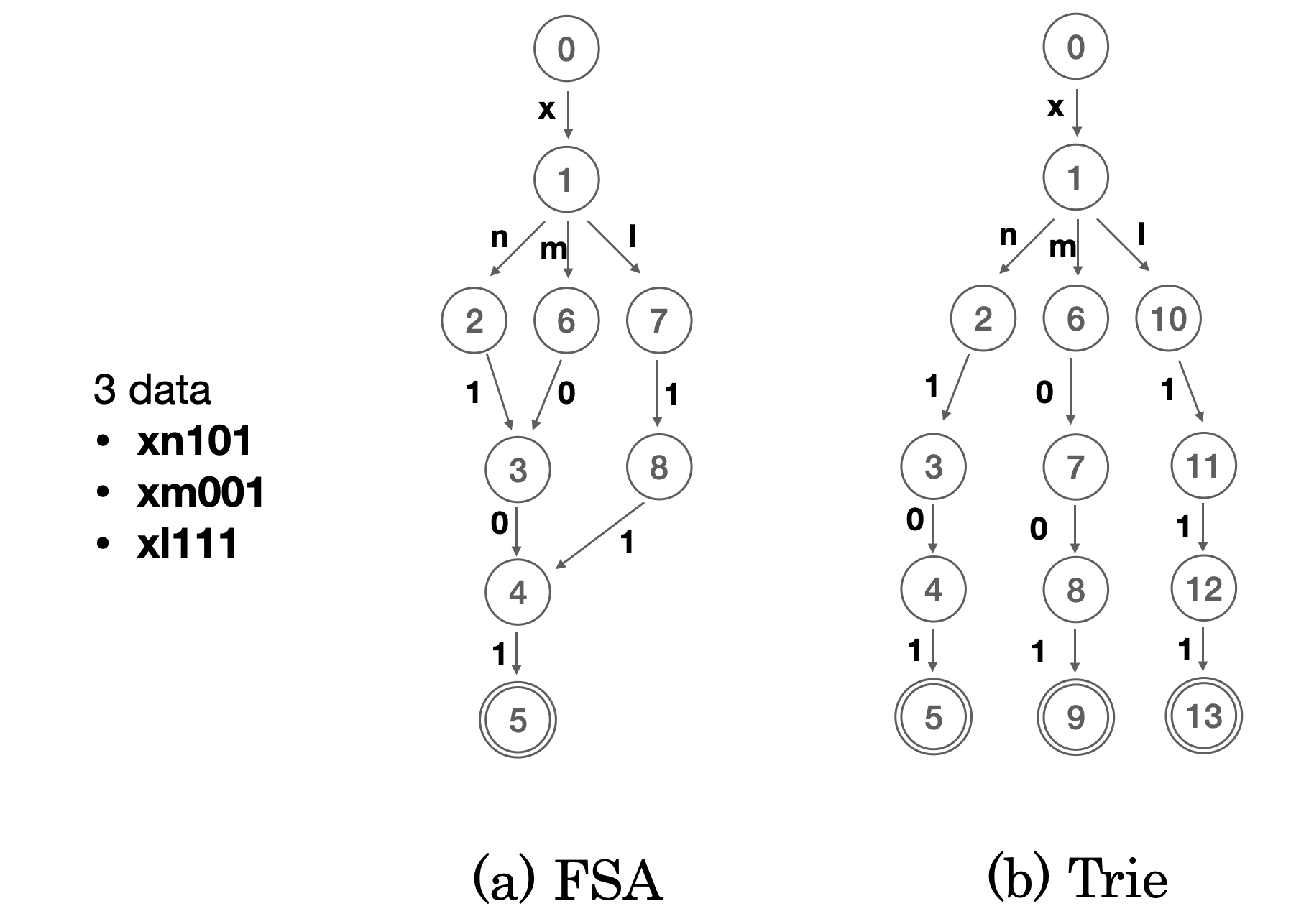}
  \caption{FSA and Trie with encoding.}
  \label{fig:fsa}
\end{figure}

Here is an example that shows how well our encoding works with FSA.
Suppose the encoded data are 'xn101', 'xm001', and 'xl111' (like short Geohash + Periodical encode). 
In that case, Trie and FSA have 8 and 13 nodes, respectively, as shown in the Figure \ref{fig:fsa}.
We can see that FSA has an advantage under the encoding described above.

\subsection{Algorithm Analysis}
\label{sec:4-4}

\begin{figure}[!t]
  \begin{algorithm}[H]
    \caption{Algorithm for PSI}
    \label{alg:pct}
    \begin{algorithmic}[1]
      \Require $q_{i}(i=1,...,n_{c})$, $\theta=(\theta_l, \theta_t)$
      \Ensure $Responses$
      \State ($R \leftarrow mapToChunkedDictionary(D, \theta)$) \algorithmiccomment{\circled{1}, \circled{2}}
      \State $loadToEnclave(q_1,...,q_{N_C})$
      \State $Q \leftarrow mapToUniqueArray(q_1,...,q_{N_C}, \theta)$ \algorithmiccomment{\circled{6}}
      \State $Results \leftarrow []$
      \For{$r_i \leftarrow R$} \algorithmiccomment{\circled{7} $R$ has $N_{D}$ chunks}
            \State $loadToEnclave(r_i)$
            \For{$value$ in $Q$} \algorithmiccomment{\circled{7} $Q$ is array with $N_{Q}$ length}
                \If{$r_i.contains(value)$}
                    \State $Results.insert(value)$
                \EndIf
            \EndFor
      \EndFor
      \State $Responses \leftarrow constructResponses(q_1,...,q_{N_C}, Results)$ \newline
      \hspace*{8em} \algorithmiccomment{\circled{8}}
      \State $Responses \leftarrow loadFromEnclave(Responses)$
    \end{algorithmic}
  \end{algorithm}
\end{figure}

Here, we discuss asymptotic computational costs of PSI and precautions.
We show our algorithm of trajectory-based PCT related to PSI part in Algorithm \ref{alg:pct}.
Some of functions are described in \ref{sec:4-1}.
Dictionary $r_i$ must implement the $contains$ method that returns a boolean value whether it includes target or not.
In the case of Hash Table, it is the computation of the hash function, and in the case of FSA, the acceptance routine with finite state automaton, both of them are asymptotically constant.
The computation costs of trajectory-based PCT is as follows.
Supposing the cost of a single key search for a dictionary is $c$ and the unique size of $Q$ is $N_{Q}$, the calculation cost is
\vspace{-3px}
\begin{equation*}
c \times N_{D}\times N_{Q} = \mathcal{O}(N_{D}N_{Q})
\vspace{-2px}
\end{equation*}
Seemingly, $N_{Q}$ and the number of chunks $N_{D}$ is constant and PSI is completely scalable for infected trajectories size.
However, note that the size of $N_{D}$ depends on the memory constraints of SGX.
When processing thousands of queries together, exact $q_i(i=1,...,N_{C})$ information needs to be kept within the enclave to correctly reconstruct the response, which can be several tens of MB in size; eventually, the size available for chunk $r_i$ is not large.
This means that actually there is a practical lower bound on $N_{D}$.

\section{Experiment and Evaluation}
\label{sec:5}
We conduct experiments using real trajectory data to demonstrate that the proposed architecture for PCT can achieve high query throughput and expected properties.

\vspace{-2px}
\subsection{Setting}
\label{sec:5-1}
\noindent
\vspace{-2px}
{\bfseries Experimental setup.}
We use an HP Z2 SFF G4 Workstation, with 4-core 3.80 GHz Intel Xeon E-2174G CPU (8 threads, with 8MB cache), 64GB RAM, and a 1TB disk, which supports SGX instruction set and has 128MB PRM (Processor Reserved Memory) in which 96MB EPC is available for use.
The host OS is Ubuntu 16.04 LTS, with Linux kernel 4.4.0-178. We use version 1.1.2 of the Rust SGX SDK\footnote[1]{https://github.com/apache/incubator-teaclave-sgx-sdk} \cite{wang2019towards} which supports Intel SGX SDK v2.9.1, and Rust nightly-2020-04-07.
Our experimental implementation and specification are available in Github\footnote[2]{https://github.com/ylab-public/PCT}.

\begin{figure*}[t]
    \begin{subfigure}[b]{0.33\textwidth}
            \includegraphics[width=0.8\linewidth]{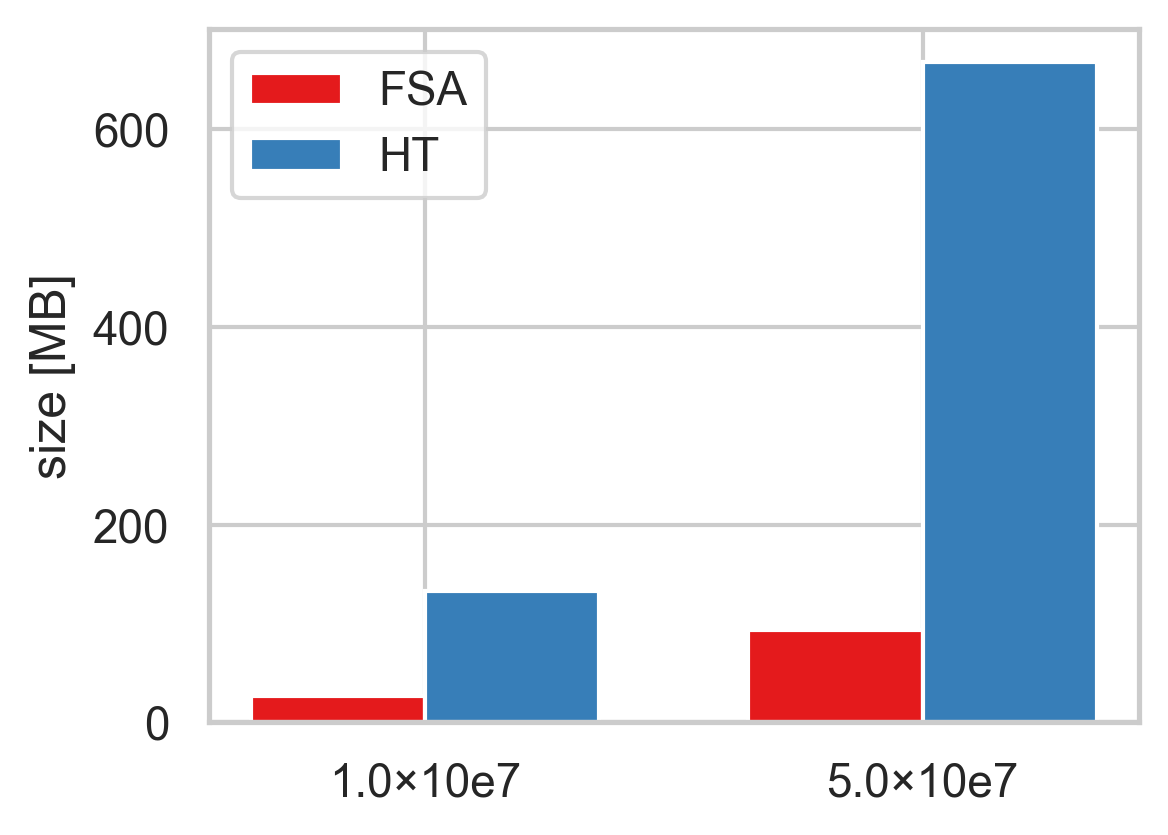}
            \caption{FSA and Hash Table.}
            \label{fig:compression1}
    \end{subfigure}%
    \begin{subfigure}[b]{0.33\textwidth}
            \includegraphics[width=0.8\linewidth]{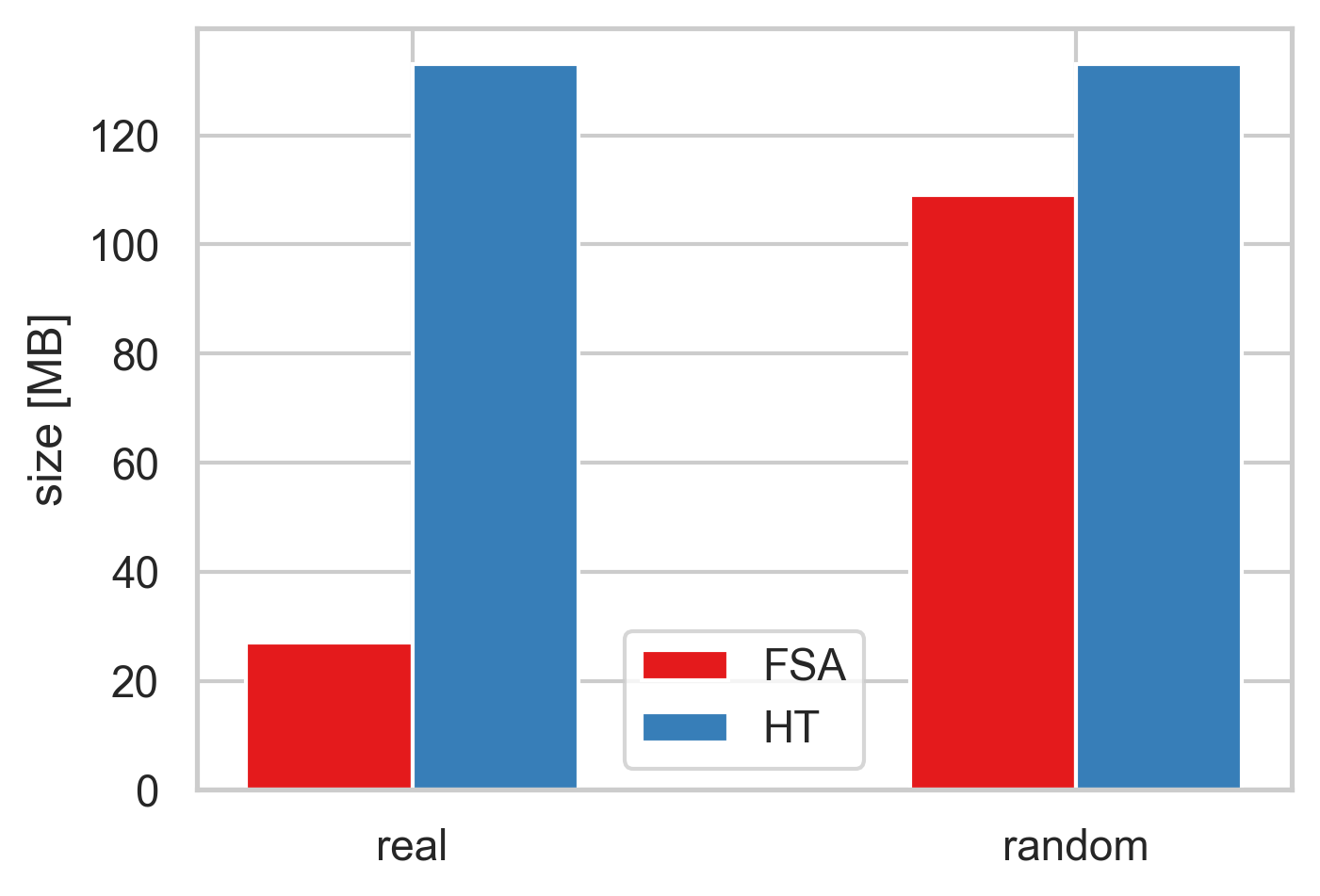}
            \caption{Real vs Random data.}
            \label{fig:compression3}
    \end{subfigure}%
    \begin{subfigure}[b]{0.33\textwidth}
            \includegraphics[width=0.8\linewidth,height=120px]{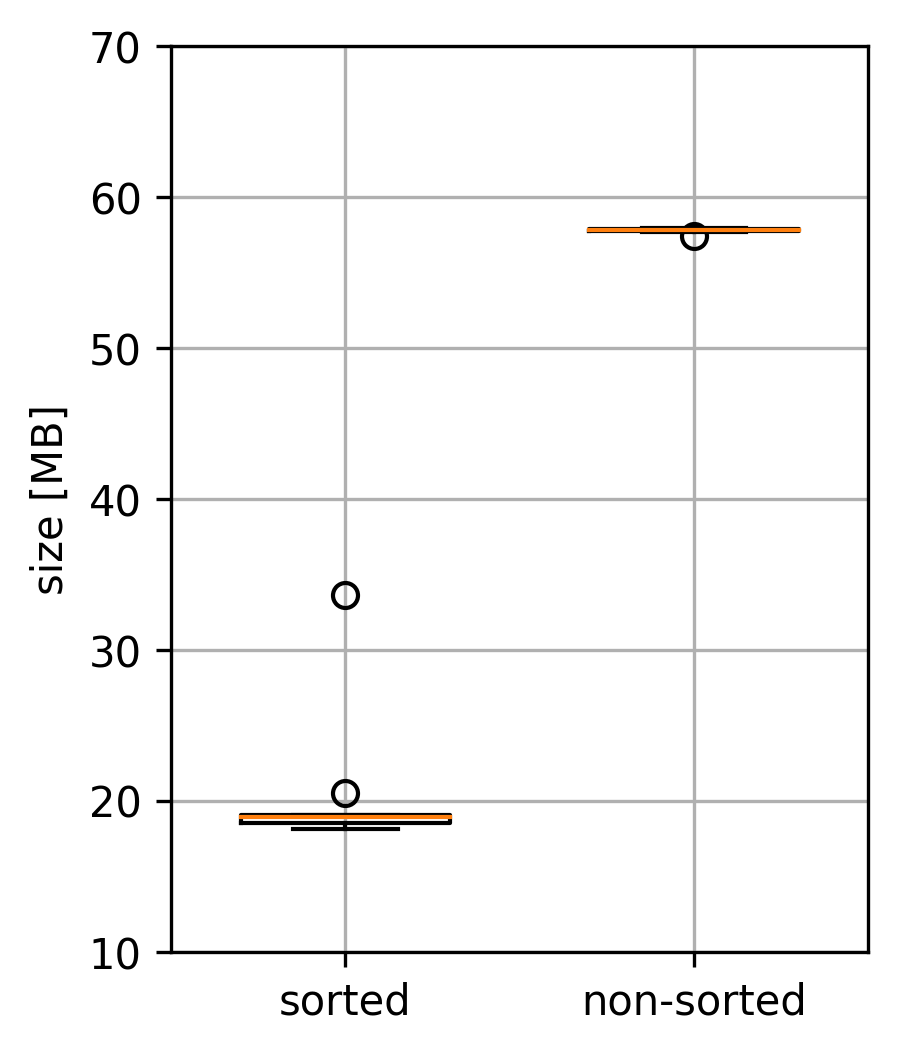}
            \caption{Sorted vs Non-sorted.}
            \label{fig:compression2}
    \end{subfigure}%
    \caption{Trajectory data compression result.}
    \label{fig:compression_results}
\end{figure*}

\noindent
{\bfseries Datasets.}
We conduct the experiments with a real dataset, including data on people's trajectories in specific regions of Japan available in JoRAS\footnote[3]{http://www.csis.u-tokyo.ac.jp} of The University of Tokyo.
We use the people flow dataset for the Kinki and Tokyo to create our experimental dataset.
We extract only the time and coordinate information and create our dataset by applying the encoding described in Section \ref{sec:4}.

\vspace{1px}
We show the appropriateness of the scale of the experiment based on actual values.
Regarding the number of data, in a practical case, considering Japan as an example, the maximum number of new infections in Japan as of July 25 was 981 per day, which means that the maximum number of new infections is approximately 14 x 1000 = 14,000 in 2 weeks.
If the trajectory data were collected every 10 minutes, the total number of trajectory data would be $2.8\times10^{7}$.
Therefore, $10^7$-$10^8$ rows of data is plausible size as the infected patient data in our experiment.
On the other hand, since the number of data for the same user available from the dataset was 1440, we adopt this as the query data for a single user.
This number corresponds the case to collect trajectories every 14 minutes over 2 weeks.

\subsection{Evaluation}
\label{sec:5-2}

\begin{table}[t]
\centering
\caption{Optimal chunk size parameters.}

\begin{tabular}{cl}
    \toprule
     & chunk size (MB)\\
    \midrule
    Hash Table & $2.0 \times 10^{6}$ (28MB) \\
    FSA & $1.0 \times 10^{7}$ (27MB) \\
    \bottomrule
\end{tabular}

\label{tab:ex_params}
\vspace{-1em}
\end{table}

First, we show the results of the data compression.
Figure~\ref{fig:compression_results} describes that our proposed method provides efficient and stable compression for real-world data.
Figure \ref{fig:compression1} shows the sizes of the $10^{7}$ and $5.0\times 10^{7}$ data stored in the FSA and the Hash Table (HT).
The size of FSA is 27MB, and HT is 133MB in $10^{7}$ data.
In addition to the small overall size, the FSA's size only increased by a factor of 3.4 while the size of the HT increased five times linearly in $5.0\times 10^{7}$. 
This indicates that the FSA can compress and hold data efficiently.
A comparison of the actual trajectory data and the randomly encoded data obtained from the uniform distribution is shown in Figure \ref{fig:compression3}.
This is the result of compression on the $10^{7}$ rows of data, which shows FSA is efficient for real trajectory data.
In relation to compression, moreover, basically it is difficult to chunk an FSA that already contains all the data.
By sorting and then chunking, similar strings will be in the same chunks, so the compression ratio can be high.
Figure \ref{fig:compression2} shows box plots of chunks of the $10^{8}$ data, sorted and then compressed, and non-sorted manner.
All the chunks achieve more efficient compression than the unsorted baseline.

\begin{figure}[t]
    \begin{subfigure}[b]{0.25\textwidth}
            \includegraphics[width=\linewidth]{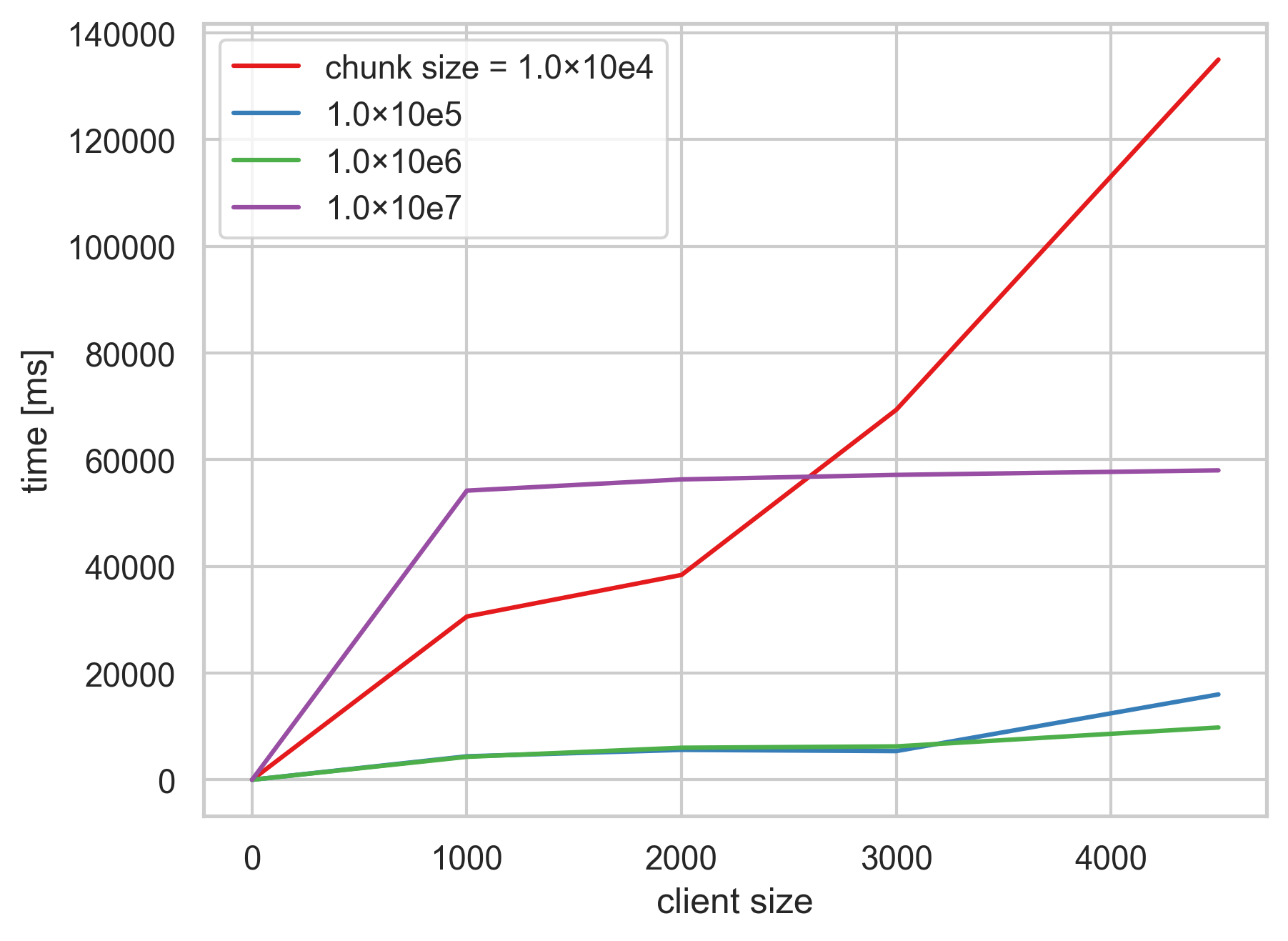}
            \caption{Hash Table chunk size.}
            \label{fig:chunk-HT}
    \end{subfigure}%
    \begin{subfigure}[b]{0.25\textwidth}
            \includegraphics[width=\linewidth]{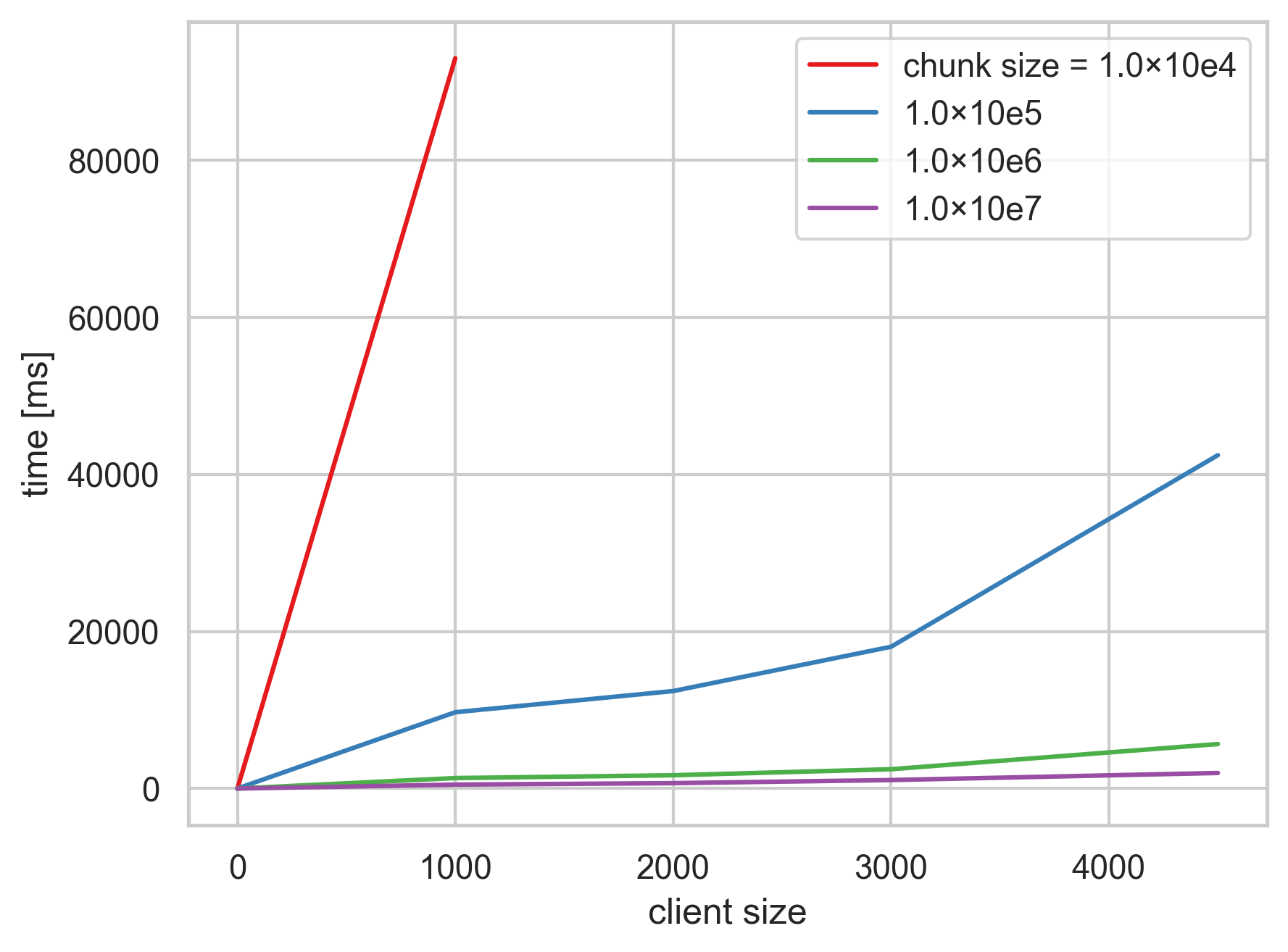}
            \caption{FSA chunk size.}
            \label{fig:chunk-FSA}
    \end{subfigure}%
    \caption{Chunk size evaluation.}\label{fig:chunk_size}
\end{figure}

\begin{figure}[t]
  \includegraphics[width=\linewidth]{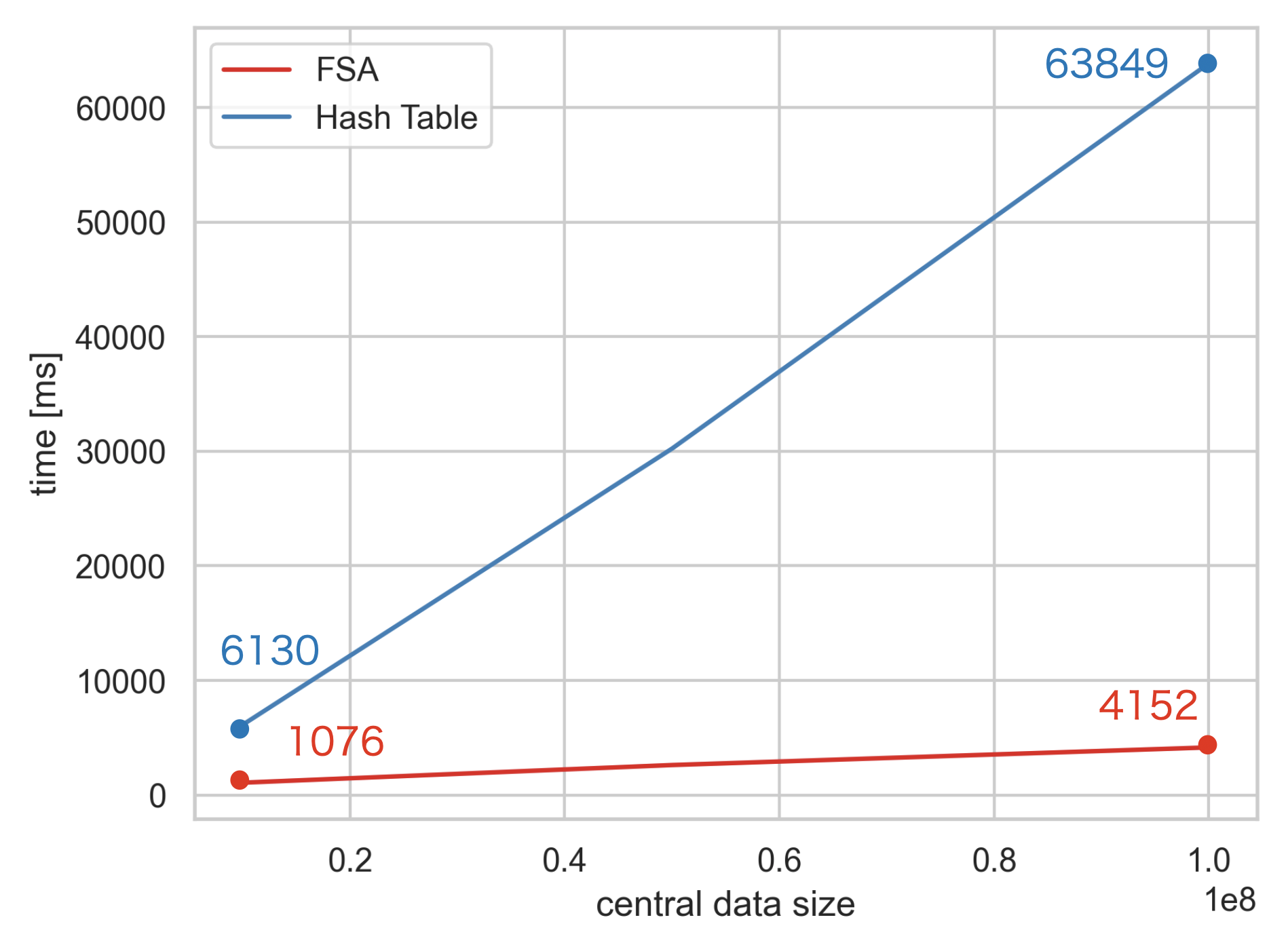}
  \caption{PSI time results.}
  \label{fig:thoroughput}
\end{figure}

\begin{figure}[t]
  \includegraphics[width=\linewidth]{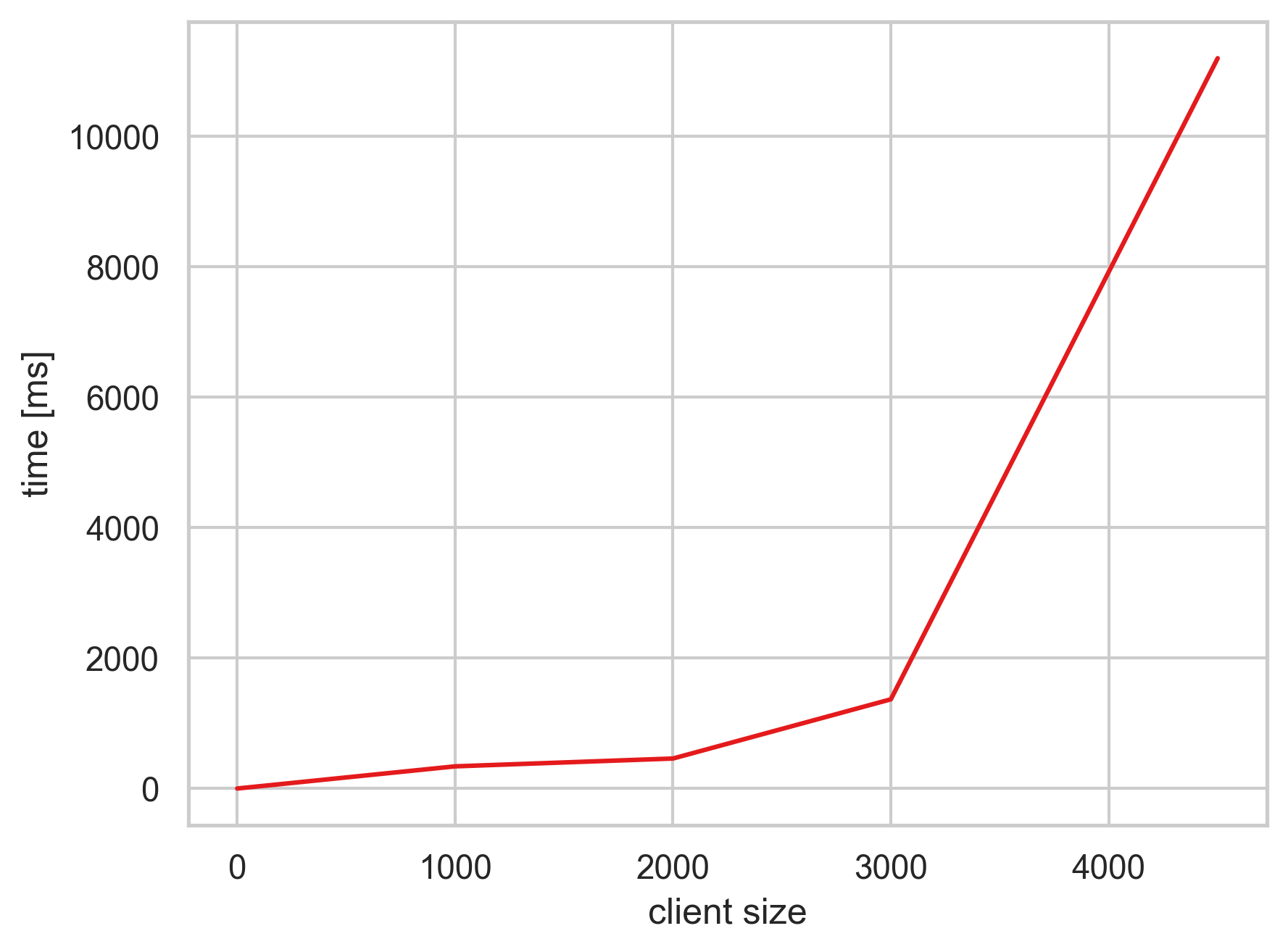}
  \caption{Assemble time in enclave for $Q$.}
  \label{fig:assemble}
\end{figure}


Next, we show the time it takes for PSI.
The appropriate chunk size differs between HT and FSA due to the different data size.
We find the best chunk size for each and use them to evaluate the execution time of PSI with the best possible performance for both of HT and FSA.
The Figure \ref{fig:chunk-HT} shows the relationship between chunk size and PSI execution time using HT for $10^{7}$ size.
From this, it can be seen that the appropriate chunk size in HT is about $10^{6}$.
This is related to the memory size of EPC, which is about 14 MB in size, a perfect match for EPC. Seemingly, about $2.0 \times 10^{6}$=28MB was the most efficient in our setting.
Similarly, Figure \ref{fig:chunk-FSA} shows in FSA, which tends to be best for FSA at around $10^{7}$.
In this case, as we saw earlier, the size is about 27 MB, which is also suitable for an EPC scale.
These are also described in \ref{tab:ex_params}.
The execution time of PSI with HT and FSA using these chunk sizes is shown in Figure \ref{fig:thoroughput}.
The maximum query throughput is 2788 when the server-side data size is $10^{7}$ and 722 when the data size is $10^{8}$.
As far as we know, there is no way so far to achieve this amount of query throughput in a trajectory-based PCT on this scale.
With $10^{7}$ data the difference between the two methods is about six times greater, and it widens to about 15 times greater as the data size increases upto $10^{8}$.
The efficiency of query processing using FSA over HT is evident.

The next thing we want to see is time-consuming process other than the PSI.
In practice, the cost of making $Q$ cannot be ignored.
The client's data $q_i$ can be only visible in enclave and merged only inside enclave, which cause a extra cost for every PCT.
The time to make $Q$ increases with the size of the client.
In the case of 3000 clients, the data size is about 60MB, and after that point, a lot of swapping starts to occur beyond the memory constraints in enclave, and then the assemble time exceeds 1 second as shown in the Figure \ref{fig:assemble}.
This noticeable result indicates that there is an appropriate query size for the amount that can be processed at one time in a single node SGX.


\section{Future Work}
\label{sec:6}



One of the future works is to further improve the security aspect.
Our current algorithm inside trusted enclave may expose data-dependent information through some side channels such as cache and/or timing and so on.
Such a {\itshape non-obliviousness} can be vulnerable even in recent TEE implementations.
A lot of attack works \cite{chen2019sgxpectre, gotzfried2017cache, infe2017} and countermeasures \cite{oblix2018, tsgx2017, krast2020efficient}.
We should study how dangerous {\itshape non-oblivious} PSI based on the trusted hardware is and how to change our algorithm to {\itshape data-oblivious} leakage with small overhead.

\section{Related work}
\label{sec:7}
\noindent
{\bfseries Privacy-preserving computation.}  
In prior private proximity testing, the most common methods proposed so far are based on advanced cryptographic approaches such as homomorphic EIGamal cryptosystem \cite{narayanan2011location}.
In contrast to these studies, we use secure hardware techniques as the starting point for protecting privacy.
As in our study, private proximity testing can be discretized and PSI as shown in \cite{narayanan2011location, lin2013private}.
While many previous studies have been focused on spatial discretization, we apply it to spatio-temporal trajectory data.
Trusted Execution Environment(TEE) techniques \cite{sabt2015trusted}, for instance Intel SGX \cite{costan2016intel}, can be promised techniques to improve such private protocols in general performance problem.
S.Tamrakar et al. \cite{tamrakar2017circle} show carousel approach in private membership test of practical malware checking scenario, which concludes that hardware-based TEE significantly accelerates their private processing.
While the overall architecture is similar to ours, two requirements of flexibility and correctness are not considered.

\noindent
{\bfseries Trajectory data representation.} 
Geohash \cite{geohash} is an encoding method from 2D latitude and longitude to string, where this property can be beneficial for certain purposes, such as fast search.
This is a public domain and probably the most pervasive geo-encoding.
Similarly, there is also recursive square segmentation encoding using quad-keys \cite{lee2011crowd}.
Our encoding partially includes these algorithms.
However, these are insufficient to represent trajectory data because they do not consider time.
In \cite{fox2013spatio}, they focus on indexing on distributed database, which is distinct from our work, but has a similar component to ours in the encoding of data.
The encoding of the time information and interleaving are different from our method since we focus on compression using FSA.

\noindent
{\bfseries Private Contact tracing.} 
There are DP3T \cite{troncoso2020decentralized} and similar schemes \cite{trieu2020epione, rivest2020pact, cen, gvili2020security} as show in section \ref{sec:1}, which are decentralized architectures using the device's wireless signals and the most popular implementation methods so far.
Reichert et al. \cite{reichert2020privacy} propose a setting similar to our study.
They also show a system for centralized contact tracing on a server using GPS data, but the proposed method in their paper is based on traditional multi party computation with ORAM, which they say underperforms in practical scenarios.
Our proposal is more practical in terms of performance and characterized by using TEE and a centralized client-server model.
As far as we know, this is the first PCT system with such features.

\section{Conclusion}
\label{sec:8}
In this paper, we proposed a trajectory-based private contact tracing system using trusted hardware to control the spread of infectious diseases.
We identified the problems of existing private contact tracing systems, clarified the requirements for trajectory-based private contact tracing, and presented an TEE-based architecture to achieve secure, efficient and flexible contact tracing.
Our experimental results with real data suggested that our proposed system can work on a realistic scale.

\section*{Acknowledgment}
This work is partially supported by JSPS KAKENHI Grant No. 17H06099, 18H04093, 19K20269, and CCF-Tencent Open Fund WeBank Special Fund.
Also, this is the result of the joint research with CSIS, the University of Tokyo (No. 974) and used the following data ID: 3000200800 (Tokyo 2008), 3038201000 (Kinki 2010).

\end{document}